\newcommand{\CLASSINPUTinnersidemargin}{1.8cm}
\newcommand{\CLASSINPUToutersidemargin}{1.8cm}
\newcommand{\CLASSINPUTtoptextmargin}{2.54cm}
\newcommand{\CLASSINPUTbottomtextmargin}{2cm}

\documentclass[letterpaper,conference]{IEEEtran}
\makeatletter
\def\ps@headings{%
\def\@oddhead{\mbox{}\scriptsize\rightmark \hfil \thepage}%
\def\@evenhead{\scriptsize\thepage \hfil \leftmark\mbox{}}%
\def\@oddfoot{}%
\def\@evenfoot{}}
\makeatother
\ifCLASSINFOpdf
\else
\fi

\usepackage[nopostdot,acronym,shortcuts,nonumberlist]{glossaries}
\makeglossaries
\newacronym{NIC}{NIC}{network interface card}
\newacronym{EMR}{EMR}{electromagnetic radiation}
\newacronym{NSA}{NSA}{National Security Agency}
\newacronym{PC}{PC}{personal computer}
\newacronym{USRP}{USRP}{Universal Software Radio Peripheral}
\newacronym{FPGA}{FPGA}{field-programmable gate array}
\newacronym{SDR}{SDR}{software-defined radio}
\newacronym{PA}{PA}{power amplifier}
\newacronym{BCI}{BCI}{bulk current injection}
\newacronym{FEC}{FEC}{forward error correction}
\newacronym{IEC}{IEC}{International Electrotechnical Commission}
\newacronym{SFD}{SFD}{start frame delimiter}
\newacronym{MAC}{MAC}{Medium Access Control}
\newacronym{DAC}{DAC}{digital-to-analog converter}
\newacronym{CSMACD}{CSMA/CD}{carrier sense multiple access with collision detection}
\newacronym{CAN}{CAN}{control area network}
\newacronym{CoE}{CoE}{CAN over Ethernet}
\newacronym{RF}{RF}{radio frequency}
\newacronym{FCS}{FCS}{frame check sum}
\newacronym{ADC}{ADC}{analog-to-digital converter}
\newacronym{(S+N)/NR}{(S+N)/NR}{signal-plus-noise-to-noise ratio}
\newacronym{ICMP}{ICMP}{Internet control message protocol}
\newacronym{DoS}{DoS}{denial of service}
\newacronym{CRT}{CRT}{cathode ray tube}
\newacronym{CTR}{CTR}{AES Counter Mode}
\newacronym{SNR}{SNR}{signal-to-noise ratio}
\newacronym{ACE}{ACE}{Authentication and Authorization for Constrained Environments}
\newacronym{RAM}{RAM}{Random Access Memory}
\newacronym{ROM}{ROM}{Read-Only Memory}
\newacronym{IoT}{IoT}{Internet of Things}
\newacronym{OAuth}{OAuth 2.0}{Open Authentication 2.0}
\newacronym{OpenID}{OpenID}{OpenID}
\newacronym{SAML}{SAML}{Security Assertion Markup Language}

\newacronym{HTTP}{HTTP}{Hypertext Transfer Protocol}
\newacronym{CoAP}{CoAP}{Constrained Application Protocol}
\newacronym{OSCORE}{OSCORE}{Object Security for Constrained RESTful Environments}
\newacronym{SSL}{SSL}{Secure Socket Layer}
\newacronym{IP}{IP}{Internet Protocol}
\newacronym{IPsec}{IPsec}{Internet Protocol Security}
\newacronym{IPv6}{IPv6}{Internet Protocol version 6}
\newacronym{6LoWPAN}{6LoWPAN}{\gls{IPv6} over Low Power Wireless Personal Area Networks}
\newacronym{LLN}{LLN}{Low Power and Lossy Networks}
\newacronym{uIP}{$\mu$IP}{$\mu$icro Internet Protocol}
\newacronym{IKE}{IKEv2}{Internet Key Exchange Protocol version 2}
\newacronym{minIKE}{minIKE}{Minimal IKE}
\newacronym{DICE}{DICE}{\gls{DTLS} for Constrained Environments}
\newacronym{ESP}{ESP}{Encapsulating Security Protocol}
\newacronym{AH}{AH}{Authentication Header Protocol}
\newacronym{DietESP}{Diet-ESP}{Diet-ESP}
\newacronym{802.15.4}{802.15.4}{IEEE 802.15.4}
\newacronym{DTLS}{DTLS}{Datagram Transport Layer Security}
\newacronym{IETF}{IETF}{Internet Engineering Task Force}
\newacronym{EDHOC}{EDHOC}{Ephemeral Diffie-Hellman Over COSE}
\newacronym{DH}{DH}{Diffie-Hellman}
\newacronym{DP}{DP}{Direct Provisioning}
\newacronym{KMP}{KMP}{Key Management Protocol}
\newacronym{PSK}{PSK}{Pre-Shared Key}
\newacronym{cert}{cert}{Certificates}
\newacronym{CPK}{CPK}{Certificate-based Public Key}
\newacronym{TCP}{TCP}{Transmission Control Protocol}
\newacronym{UDP}{UDP}{User Datagram Protocol}
\newacronym{IEEE}{IEEE}{Institute of Electrical and Electronics Engineers}
\newacronym{OSI}{OSI}{Open Systems Interconnection}
\newacronym{PHY}{PHY}{Physical Layer}
\newacronym{ISM}{ISM}{Industrial Scientific Medical}
\newacronym{DSSS}{DSSS}{Direct Sequence Spread Spectrum}
\newacronym{CSMACA}{CSMA/CA}{Carrier Sense Multiple Access with Collision Avoidance}
\newacronym{REST}{REST}{Representational State Transfer}
\newacronym{AES}{AES}{Advanced Encryption Standard}
\newacronym{COSE}{COSE}{CBOR Object Signing and Encryption}
\newacronym{CBOR}{CBOR}{Concise Binary Object Representation}
\newacronym{MSK}{MSK}{Minimum-Shift Keying}
\newacronym{URI}{URI}{Universal Resource Identifier}
\newacronym{RPL}{RPL}{Routing Protocol for Low-Power and Lossy Networks}
\newacronym{AS}{AS}{Authorization Server}
\newacronym{RS}{RS}{Resource Server}
\newacronym{RO}{RO}{Resource Owner}
\newacronym{C}{Client}{Client}
\newacronym{PoP}{PoP}{proof-of-possession}
\newacronym{PoPK}{PoPK}{PoP Key}
\newacronym{TTP}{TTP}{Trusted Third Party}
\newacronym{SA}{SA}{Security Association}
\newacronym{JSON}{JSON}{JavaScript Object Notation}
\newacronym{SPI}{SPI}{Security Parameters Index}
\newacronym{ISAKMP}{ISAKMP}{Internet Security Association Key Management Protocol}
\newacronym{TLS}{TLS}{Transport Layer Security}
\newacronym{RPK}{RPK}{Raw Public Key}
\newacronym{SIGMA}{SIGMA}{SIGn-and-MAc}
\newacronym{cnf}{\emph{cnf}}{confirmation}
\newacronym{SoC}{SoC}{System-On-Chip}
\newacronym{OS}{OS}{Operating System}
\newacronym{AEAD}{AEAD}{Authenticated Encryption and Associated Data}
\newacronym{IV}{IV}{Initialization Vector}
\newacronym{HMAC}{HMAC}{Hash-based MAC}
\newacronym{PRF}{PRF}{Pseudo-Random Function}
\newacronym{RDC}{RDC}{Radio Duty-cycle}
\newacronym{DAG}{DAG}{Direct Acyclic Graph}
\newacronym{VPN}{VPN}{Virtual Private Network}
\newacronym{CORE}{CORE}{Constrained RESTful Environments }

\newacronym{SICS}{RISE-SICS}{Research Institutes of Sweden: Swedish Institute of Computer Science}

 \usepackage{graphicx}
\usepackage{hyperref}
\usepackage{multirow}
\usepackage{amssymb}
\usepackage{rotating}
\usepackage{subcaption}
\usepackage{tikz}
\newcommand\copyrighttext{%
  \footnotesize \copyright{} 2018 IEEE. Personal use of this material is permitted. Permission from IEEE must be obtained for all other uses, in any current or future media, including reprinting/republishing this material for advertising or promotional purposes, creating new collective works, for resale or redistribution to servers or lists, or reuse of any copyrighted component of this work in other works. The official version can be found at \url{http://dx.doi.org/10.1109/CNS.2018.8433209}}
  
\newcommand\copyrightnotice{%
\begin{tikzpicture}[remember picture,overlay]
\node[anchor=south,yshift=10pt] at (current page.south) {\fbox{\parbox{\dimexpr\textwidth-\fboxsep-\fboxrule\relax}{\copyrighttext}}};
\end{tikzpicture}%
}

\begin{document}

\title{ACE of Spades in the IoT Security Game: \\A Flexible IPsec Security Profile for Access Control}

\newcommand{\fmsr}[1]{{\textbf{[SR!]}}\footnote{{\textbf{[Shahid]: }}#1}}
\newcommand{\fmzh}[1]{{\textbf{[MT!]}}\footnote{{\textbf{[Marco]: }}#1}}
\author{\IEEEauthorblockN{Santiago Aragon\IEEEauthorrefmark{1}\IEEEauthorrefmark{2},
Marco Tiloca\IEEEauthorrefmark{2},
Max Maass\IEEEauthorrefmark{1},
Matthias Hollick\IEEEauthorrefmark{1} and
Shahid Raza\IEEEauthorrefmark{2}
}
\IEEEauthorblockA{\IEEEauthorrefmark{1}Technische Universit\"at Darmstadt, Secure Mobile Networking Lab\\
Mornewegstr. 32, 64293 Darmstadt, Germany\\
\{saragon, mmaass, mhollick\}@seemoo.tu-darmstadt.de}
\IEEEauthorblockA{\IEEEauthorrefmark{2}RISE SICS, Security Lab\\
Isafjordsgatan 22, SE-16440 Kista, Sweden \\\{marco.tiloca, shahid.raza\}@ri.se}}

\maketitle

\copyrightnotice

\begin{abstract}
The Authentication and Authorization for Constrained Environments (ACE) framework provides fine-grained access control in the Internet of Things, where devices are resource-constrained and with limited connectivity. The ACE framework defines separate profiles to specify how exactly entities interact and what security and communication protocols to use. This paper presents the novel ACE IPsec profile, which specifies how a client establishes a secure IPsec channel with a resource server, contextually using the ACE framework to enforce authorized access to remote resources. The profile makes it possible to establish IPsec Security Associations, either through their direct provisioning or through the standard IKEv2 protocol. We provide the first Open Source implementation of the ACE IPsec profile for the Contiki OS and test it on the resource-constrained Zolertia Firefly platform. Our experimental performance evaluation confirms that the IPsec profile and its operating modes are affordable and deployable also on constrained IoT platforms.\end{abstract} 

\IEEEpeerreviewmaketitle

\section{Introduction}
\label{s:introduction}

The Internet of Things (IoT) refers to network scenarios where billions of devices communicate over IP networks and are available on the Internet. This includes everyday objects and appliances, and has been constantly fostering a number of use cases and business opportunities, from sensor and actuator networks to smart buildings, from monitoring of critical infrastructures to controlled resource sharing. As more and more applications are being developed, the IoT is expected to have a huge impact on the way we live and work.

At the same time, security plays a fundamental role, even during this transition process. In fact, ensuring security in IoT scenarios is of vital importance to counteract information breaches and service dysfunctions, which may result in severe performance degradation and privacy violations, or even threaten safety of people and infrastructures. Securing the IoT is thus vital to ensure its successful deployment and adoption.

However, unlike in traditional networks, IoT devices are typically resource-constrained, i.e. equipped with limited resources. That is, they are scarce as to processing power, storage and energy availability, often being battery-powered. Besides, most IoT devices are wirelessly connected over low-power and lossy networks, thus exhibiting limited connectivity and availability. Also, they often lack traditional user interfaces, and are likely deployed in unattended environments. As a result, protecting billions of IoT devices with traditional approaches is challenging, which fosters the development of novel security solutions suitable for the IoT. Yet, many of these solutions do not base on established standards and are difficult to scrutinize in terms of their security guarantees. 
\begin{figure}[t]
\centering
\includegraphics[width=.5\columnwidth]{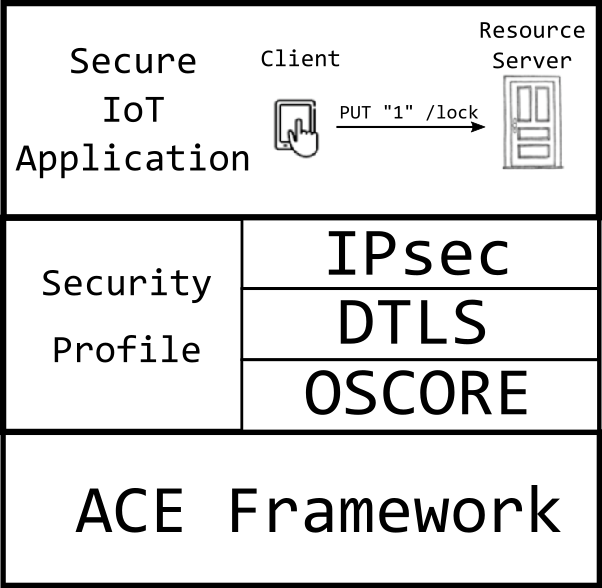}
\caption{IoT applications can secure their communications by using the ACE framework and the different security profiles.}
\label{fig:ace_profiles-stack}
\end{figure}

The first security challenge consists in efficiently enabling secure communication and message exchange. Due to the resource-constrained nature of typical IoT devices, their great heterogeneity and their large-scale deployment, it is not feasible to rely on solutions for traditional network environments. To this end, a number of secure communication protocols for the IoT are available and have been increasingly adopted in constrained environments. In particular, \cite{raza13lithe} and \cite{Raza2011SecuringIO} show how 6LoWPAN header compression mechanisms optimize security protocols to be deployable in resource-constrained networked scenarios. However, it can be very difficult to provision millions, or even billions, of resource-constrained IoT devices with the cryptographic keys necessary to securely communicate and operate. Even the establishment of secure sessions based on pre-shared symmetric keys can easily result in hard-to-manage and poorly scalable key distribution.

The second critical security aspect concerns authorization and access control. Typically, a \emph{Client} wants to access a resource hosted on a \emph{Resource Server} (RS), which is often deployed as a resource-constrained device. This requires the Client and the RS to mutually authenticate, and must permit the RS to verify Client requests as previously authorized. In order to enable fine-grained and flexible access control in the IoT, the \emph{Authentication and Authorization for Constrained Environments} (ACE) framework has been proposed \cite{ietf-ace-oauth-authz-07}, building on the authorization framework OAuth 2.0 \cite{rfc6749}.

The ACE framework relies on an \emph{Authorization Server} (AS), that has a trust relation with the RS and authorizes resource accesses from requesting Clients, based on pre-defined policies. However, the ACE framework admits the definition of separate \emph{profiles} describing how these actors interact with each other and what communication and security protocols they use. A few profiles have been proposed, including \cite{ietf-ace-dtls-authorize-02} for the DTLS protocol \cite{rfc6347}, as well as \cite{ietf-ietf-oscore-profile-00} for OSCORE \cite{ietf-core-object-security-07}. The choice of the particular profile to use has to take into account the specific use case and its security requirements, as well as the related trust and security models. This naturally leads to the most suitable communication and security protocols to adopt, and hence to the related profile describing how to use them in the ACE framework. Figure \ref{fig:ace_profiles-stack} shows how an IoT application for home automation can leverage on the ACE Framework profiles, e.g. a traditional Internet host like a Smart phone can secure its communications with a smart lock using IPsec, DTLS or OSCORE.

This paper presents the novel ACE IPsec profile, which describes how Client and RS set up and use an IPsec channel \cite{rfc4301}, contextually with the access control enforced by the AS. The profile displays two key benefits tightly paired with the access control provided by the ACE framework.

First, it enables secure communication between Client and RS at the network layer, by flexibly leveraging the IPsec security protocols AH \cite{rfc4302} and ESP \cite{rfc4303}, and thus counteracting network-layer attacks such as IP spoofing. This is fundamentally achieved by establishing IPsec \emph{Security Associations} between Client and RS. Second, it efficiently addresses the provisioning of key material, by embedding the process in the authorization workflow of the ACE framework and taking advantage of the AS. Specifically, the IPsec Security Associations can be generated by the AS, and then directly provided to Client and RS. As an alternative, the AS provides the Client and RS with the necessary key material to establish the IPsec Security Associations through the standard IKEv2 protocol \cite{rfc7296}, based either on symmetric or asymmetric cryptography.

In order to encourage wider acceptance and interoperability across multiple vendors, we submitted a draft description of our profile to the IETF for possible standardization \cite{aragon-ace-ipsec-profile-01}. The draft focuses on the theoretical contribution and practical considerations, and it does not refer to a particular implementation or experimental evaluation of the IPsec profile.

In this paper, we additionally describe our implementation of the ACE framework and the ACE IPsec profile for the Contiki OS \cite{contiki}. We test it on real IoT devices using the resource-constrained Zolertia Firefly platform \cite{zolertia}. Our implementation covers all the actors in the ACE framework and is available as open source software at \cite{IPsec_profile_impl}. To the best of our knowledge, this is the first implementation of the ACE framework for the Contiki OS, and the first one ever of its IPsec profile. Additionally, it targets scenarios where even the AS is a resource-constrained device.

We utilize our implementation to experimentally evaluate the performance of the ACE framework when using the novel IPsec profile under different channel establishment and authentication methods. In particular, we consider message size, memory and energy consumption, and time required for the Client to perform an authorized resource access at the RS. Our results confirm that the IPsec profile is affordable on resource-constrained devices, and hence is effectively deployable in IoT scenarios to enforce access control paired with IPsec-based secure communication.

The rest of the paper is organized as follows. Section \ref{s:related_work} discusses the related work. Section \ref{s:background} introduces background concepts and technologies. In Section \ref{s:protocol_overview}, the ACE IPsec profile is introduced. Section \ref{sec:performance_evaluation} presents our performance evaluation. Finally, Section \ref{s:conclusion} draws conclusive remarks.

\section{Related Work}
\label{s:related_work}

Different profiles have been proposed for the \gls{ACE} framework. In \cite{ietf-ace-dtls-authorize-02}, Gerdes \textit{et al.} describe the \gls{DTLS} profile, which delegates the authorization and authentication of a Client device to the establishment of a \gls{DTLS} session \cite{rfc6347} between the Client and a \gls{RS}. Specifically, \gls{DTLS} can be used in the symmetric \gls{PSK} mode or the asymmetric \gls{RPK} mode. If the \gls{PSK} mode is used, the successful establishment of a \gls{DTLS} session also acts as a \gls{PoP} for the Client's \gls{PSK}. In case the RPK mode is used, the Client is authenticated through its asymmetric public key. Finally, this profile uses the \gls{CoAP} \cite{rfc7252} over \gls{DTLS} between Client and \gls{RS}. The feasibility of securing \gls{CoAP} messages with \gls{DTLS} has been investigated in \cite{raza13lithe}. 

The OSCORE profile of ACE proposed by Seitz \emph{et al.} \cite{ietf-ietf-oscore-profile-00} provides communication security between Client and RS by means of the \gls{OSCORE} protocol \cite{ietf-core-object-security-07}. OSCORE ensures request/response binding and selectively protects CoAP messages at the application layer, by using the compact \emph{CBOR Object Signing and Encryption} (COSE) \cite{rfc8152} based on the \emph{Concise Binary Object Representation} (CBOR) \cite{rfc7049} as data encoding format. This provides true \emph{end-to-end} secure communication between Client and RS, even in the presence of (untrusted) intermediary CoAP proxies, which remain able to perform their intended operations (e.g. message caching). This is not possible when \gls{DTLS} is used, as it requires transport-layer security to be terminated at the proxy, which is thus able to inspect and possibly alter the entire content of CoAP messages exchanged between Client and RS.  A secure context can be established directly from a symmetric PoP key, or by using external key establishment protocols. Currently, the DTLS and OSCORE profiles have not been implemented or evaluated for resource-constrained IoT devices.

Compared to the OSCORE profile, the IPSec profile presented in this paper preserves and leverages a flexible key establishment based on the IKEv2 protocol \cite{rfc7296}, tightly paired with ACE authorization process. In addition, it makes it possible to employ policy-based traffic filtering, also during the actual establishment of IPsec channels between Client and RS. In contrast, this feature is not available for the \gls{DTLS} and \gls{OSCORE} profiles. Besides, we have implemented the IPsec profile together with the ACE framework on the Contiki OS, and tested it over resource-constrained IoT devices.

Finally, Sciancalepore \emph{et al.} propose a different authorization framework for the IoT \cite{Sciancalepore2017}, also based on OAuth 2.0 and other standard protocols. In particular, it provides access control through an intermediary gateway acting as mediator between IoT networks and non-constrained Internet segments. However, unlike the ACE framework, \cite{Sciancalepore2017} displays a considerably higher level of complexity and requires the intermediary gateway to be fully trusted. %
\section{Background}
\label{s:background}

In this section, we review the main concepts and building blocks considered by the IPsec profile presented in the paper.

\subsection{OAuth 2.0}
\label{ss:oauth}

A typical security requirement in the Internet is \emph{authorization}, i.e. the process for granting approval to a \emph{client} that wants to access a resource \cite{rfc4949}. The \gls{OAuth} authorization framework has asserted itself among the most adopted standards to enforce authorization \cite{rfc6749}. \gls{OAuth} relies on an \glsdisp{AS}{\emph{Authorization Server} (AS)} entity, and addresses all common issues of alternative approaches based on credential sharing, by introducing a proper authorization layer and separating the role of the actual \emph{resource owner} from the role of the \emph{client} accessing a resource.

Specifically, \gls{OAuth} allows a client entity (e.g. a user, a host) to obtain a specific and limited access to a remote resource, hosted at a \emph{Resource Server} (\gls{RS}), while enforcing the permission from the original resource owner. That is, the resource owner grants authorization through the intermediary AS, which in turn provides the client with an \emph{access token} including the actual authorization information. Access tokens consist of strings that are opaque to the client and encode decisions for authorized resource access in terms of duration and scope. Such decisions are ultimately taken by the AS and enforced by the RS upon processing the access token.

In addition, the AS prevents non-authorized parties from tampering with issued access token or possibly generating bogus ones. To this end, the client presents the access token to the RS upon accessing the intended resource. Then, the RS verifies that the access token is valid, before proceeding with processing and serving the request from the client. This requires that: i) the client is pre-registered at the AS; ii) the AS securely communicates with both the client and the RS; and iii) the AS and RS have pre-established a trust relation. An AS may be associated with multiple RSs at the same time. The involved parties perform RESTful interactions via the HTTP protocol \cite{rfc7231}, contacting the RESTful \emph{endpoints} associated to specific steps in the \gls{OAuth} flow.

The approach adopted by \gls{OAuth} has become more and more important in IoT scenarios, where heterogeneous and resource constrained devices are deployed on a large scale, often configured as RS. However, these peculiarities make \gls{OAuth} \emph{as is} not suitable for the IoT. This motivated the design of the \emph{Authentication and Authorization for Constrained Environments} (\gls{ACE}) framework \cite{ietf-ace-oauth-authz-07}, as a standard proposal under the Internet Engineering Task Force (IETF). The \gls{ACE} framework builds on \gls{OAuth} in order to adapt and extend it for enforcing authorization in constrained IoT environments. To this end, it uses the basic \gls{OAuth} mechanisms where possible, while also providing application developers with extensions, profiles and additional guidance to ensure a privacy-oriented and secure usage.

\paragraph{Actors} The \gls{ACE} framework considers the following four actors, in accordance with the main paradigm inherited from \gls{OAuth}.

\noindent
\textbf{Client}: the entity accessing a remote protected resource.\\
\noindent
\textbf{Resource Server} (\gls{RS}): the entity hosting protected resources and serving requests from authorized clients. Authorization is enforced through access tokens that requesting clients provide to the endpoint \emph{/authz-info} at the RS via a POST request.\\
\noindent
\textbf{Authorization Server} (\gls{AS}): the entity authorizing Clients to access protected resources at the \gls{RS}. The \gls{AS} is typically equipped with plenty of resources and hosts two endpoints: i) the \emph{/token} endpoint, for receiving Access Token Requests from Clients; and ii) the \emph{/introspect} endpoint, that the \gls{RS} can use to query for extra information on received access tokens.\\
\noindent
\textbf{Resource Owner} (RO): the entity owning a protected resource hosted at the \gls{RS}, and entitled to grant access to it. The RO can dynamically provide its consent for giving a Client access to a protected resource, according to the traditional OAuth flows. However, the ACE framework is especially tailored to resource-constrained settings, where such consent is typically pre-configured as authorization policies at the \gls{AS}. Such policies are then evaluated by the \gls{AS} upon receiving a token request from a Client. In particular, the policies from the RO influence what claims the \gls{AS} ultimately includes into the access token released to a requesting client.

\paragraph{Building Blocks} From an operational point of view, the \gls{ACE} framework consists of the following building blocks.

\noindent
\textbf{OAuth} \cite{rfc6749} defines the overall authentication paradigm resulting in the protocol flows and actors' interaction.

\noindent
\textbf{CoAP} \cite{rfc7252} is a RESTful application-layer protocol for the IoT, typically running over UDP and able to greatly limit overhead and message exchanges. As \gls{CoAP} is lightweight and tailored to resource-constrained IoT devices, it is the preferred choice in the \gls{ACE} framework. Also, \gls{CoAP} has been designed to explicitly support operations of intermediary Proxy nodes.

\noindent
\textbf{\gls{CBOR}} \cite{rfc7049} is a compact version of \gls{JSON}~\cite{rfc7159}, i.e. a light-weight format for data interchange which is easy to create and process. In particular, \gls{CBOR} enables binary encoding of small messages conveying self-contained access tokens, \gls{CoAP} POST parameters, and \gls{CoAP} responses.

\noindent
\textbf{\gls{COSE}} \cite{rfc8152} enables application-layer security in the \gls{ACE} framework, especially in order to secure access tokens.

\paragraph{Authorization credentials} In order to access protected resources hosted at a \gls{RS}, a Client must get the right authorization credentials in the form of an access token. Specifically, an access token is a data structure including authorization permissions issued by the \gls{AS}, provided to the Client, and delivered to the \gls{RS} for authorized resource access. Access tokens are opaque to the Client, i.e. their semantics are unknown to the Client, and are cryptographically protected, e.g. by means of COSE \cite{rfc8152}. That is, access tokens are intelligible only to the \gls{RS} and the \gls{AS}.

A proof-of-possession (PoP) token is an access token bound to a cryptographic key, which is used by the \gls{RS} to authenticate a Client request. \gls{PoP} tokens rely on the \gls{AS} to act as \gls{TTP}, in order to bind a PoP key (PoPK) to an access token. PoP keys can be based on symmetric or asymmetric cryptography. In case of a symmetric PoP key, the \gls{AS} generates it and provides it to the Client and the \gls{RS}. To this end, the \gls{AS} can: i) make it available at the \emph{/introspect} endpoint; or ii) provide it to the \gls{RS} (Client) in the access token (Access Token Response). For asymmetric PoP keys, the Client generates a key pair, and provides the public key to the \gls{AS} in the Access Token Request. Also, the \gls{AS} provides the \gls{RS}' public key to the Client in the Access Token Response. The Client's public key is made available to the \gls{RS} through the \emph{/introspect} endpoint, or conveyed in the access token.

The \gls{ACE} framework delegates to separate \emph{security profiles} the description of how enforcing secure communication and mutual authentication among the involved parties, as well as the details about their specific interactions. In particular, a security profile must specify: i) the communication and security protocols between the \gls{RS} and the Client, as well as the methods to achieve mutual authentication; ii) the communication and security protocols for interactions between the Client and the \gls{AS}; iii) the \gls{PoP} protocols to use and how to select one; and iv) the mechanisms to protect the \emph{/authz-info} endpoint at the \gls{RS}. The \gls{AS} informs the Client of the specific profile to use by means of the \emph{profile} parameter in the Access Token Response. Also, the \gls{AS} is expected to know what profiles are supported by the Client and \gls{RS}.
\begin{figure}[t]
\centering
\includegraphics[width=0.88\columnwidth]{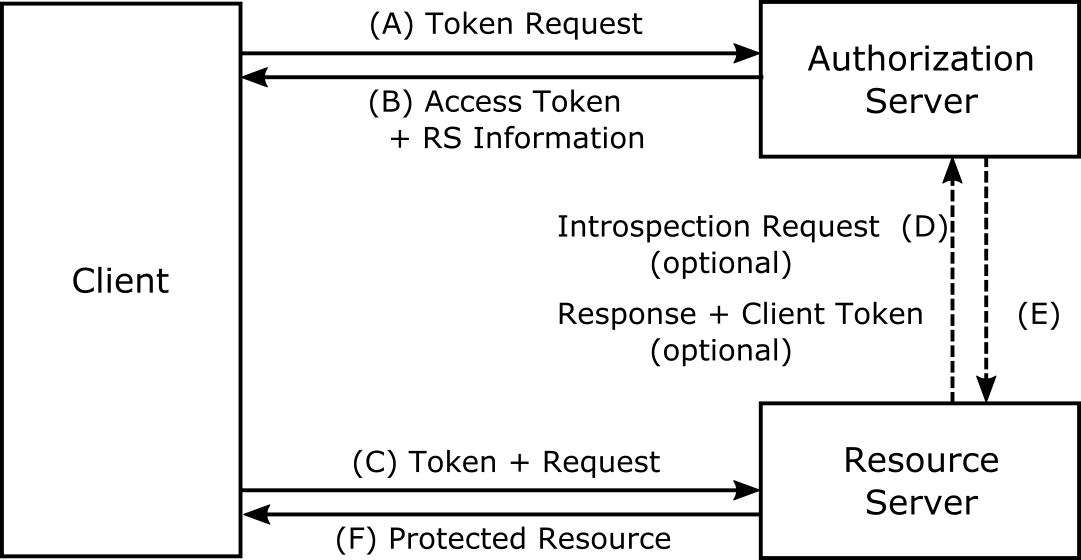}
\caption{\gls{ACE} framework's protocol flow (adapted from \cite{ietf-ace-oauth-authz-07}).}
\label{fig:ace_prot}
\end{figure}
\subsection{The ACE framework}
\label{ss:ace_framework}

The protocol flow in the ACE framework consists of the following steps, also shown in Figure \ref{fig:ace_prot}. Communications between Client and \gls{AS} as well as between \gls{RS} and \gls{AS} should be secured, in accordance with the used security profile.\\
\noindent
\textbf{(A) Access Token Request}. The Client sends an \emph{Access Token Request} to the \emph{/token} endpoint at the \gls{AS}. This includes information about the Client's credentials and the requested permissions for accessing the protected resource at the \gls{RS}.\\
\noindent
\textbf{(B) Access Token Response + RS Information}. Once it has successfully processed the Access Token Request received at step (A), the \gls{AS} generates an \emph{access token} and provides it to the Client in the Access Token Response. The access token and the Access Token Response include also \gls{RS} information and relevant parameters such as token type, expiration time, scope, state, security profile to be used and \gls{PoP} key.\\
\noindent
\textbf{(C) Token + Request}. First, the Client sends the access token to the \emph{/authz-info} endpoint at the \gls{RS}, followed by the actual Resource Request at the specific resource endpoint. The \gls{RS} and the Client authenticate each other and set up a secure communication channel, according to the security profile specified by the AS and the related \gls{PoP} keys. The \gls{RS} can validate the access token entirely by itself, or by interacting with the \gls{AS} via the \emph{/introspect} endpoint.\\
\noindent
\textbf{(D) Introspection Request}. The \gls{RS} may send the access token for validation to the \emph{/introspect} endpoint at the \gls{AS}. If the access token is self-contained and the \gls{RS} can validate it by itself, this step as well as step (E) can be omitted.\\
\noindent
\textbf{(E) Token Introspection Response + Client Token}. Upon receiving an Introspection Request from the \gls{RS}, the \gls{AS} validates the access token and replies to the \gls{RS}. The response includes extra information to achieve mutual authentication between the Client and the \gls{RS}. Additional parameters meant to be forwarded to the Client are sent as a Client Token.\\
\noindent
\textbf{(F) Protected Resource Response}. Once the access token has been successfully validated and a secure channel has been established, the \gls{RS} processes the Resource Request received from the Client at step (C). Then, the \gls{RS} provides the requested resource to the Client over the established secure channel, in accordance with the used security profile.

\subsection{IPsec and IKEv2}
\label{ss:IPsec_and_IKEv2} 
The IPsec suite is a collection of protocols to secure IP-based communications at the network layer \cite{rfc4301}. It fundamentally relies on \glsdisp{SA}{\emph{Security Associations} (SAs)}, each of which describes how to secure a one-way channel between two parties. Thus, two SAs are required to secure a two-way communication channel. An IPsec \gls{SA} is identified by a \gls{SPI}, and it specifies cryptographic material, as well as the parameters and protocols to secure IP packets through the IPsec channel. This includes the security protocol to be used, i.e. \gls{AH} \cite{rfc4302} or \gls{ESP} \cite{rfc4303}.

In particular, \gls{AH} enables connectionless integrity and data origin authentication. Instead, \gls{ESP} provides confidentiality, data origin authentication, connectionless integrity, replay protection and limited traffic flow confidentiality. Although both protocols provide integrity protection, \gls{AH} additionally protects the header of IP packets. Both \gls{AH} and \gls{ESP} can operate in two modes, namely \emph{transport} and \emph{tunnel}. The former processes IP packets without changing the IP headers, while the latter encapsulates the original IP packet into a new one, thus protecting its payload and header.

Finally, \glspl{SA} are established manually or dynamically, e.g. by using \gls{IKE} \cite{rfc7296} as key exchange protocol. In particular, \gls{IKE} enables mutual authentication between two parties through a \gls{DH} key exchange, using the pre-shared key (PSK) or the certificate raw public key (Cert) mode. The usual execution of \gls{IKE} consists of two pairs of request/response messages, i.e. IKE\_SA\_INIT and IKE\_AUTH. This establishes: i) an \gls{IKE} \gls{SA} to protect \gls{IKE} traffic; and ii) a first IPsec \gls{SA} to protect the actual IP traffic. Further SAs can be derived through CREATE\_CHILD\_SA messages. %
\section{Protocol Overview}
\label{s:protocol_overview}

In this section, we describe the ACE IPsec profile. The profile provides an operative instance of the \gls{ACE} framework, by defining the communication and security protocols used by a Client to perform an authenticated and authorized access to a protected resource hosted at a Resource Server (RS). In particular, it considers the IPsec protocol suite and the \gls{IKE} key management protocol to enforce secure communications between Client and \gls{RS}, server authentication and proof-of-possession bound to an ACE access token.

Hereafter, we denote with SA-C the \gls{SA} used for the unidirectional IPsec channel from the Client to the \gls{RS}, while with SA-RS the \gls{SA} used for the unidirectional IPsec channel from the \gls{RS} to the Client. Also, information to build \gls{SA}s is encoded as the newly introduced \emph{ipsec} structure in the ACE access token. Such information includes: i) two \glspl{SPI}, namely \emph{SPI\_SA\_C} and \emph{SPI\_SA\_RS}; ii) the IPsec mode, i.e. transport or tunnel; iii) the security protocol, i.e. \gls{AH} or \gls{ESP}; iv) cryptographic keys; v) the key establishment method to fully setup the two-way IPsec channel; and vi) the SAs' lifetime. In particular, \emph{SPI\_SA\_C} (\emph{SPI\_SA\_RS}) refers to SA-C (SA-RS). In case tunnel mode is chosen, source and destination IP addresses are also specified.

\subsection{Key Establishment Methods}\label{subsec:methods} 
The IPsec profile provides three methods for establishing a pair of \gls{SA}s, and hence a two-way IPsec channel between Client and RS. The three methods are: i) \emph{Direct Provisioning} (DP); ii) establishment with IKEv2 and symmetric-key authentication; and iii) establishment with IKEv2 and asymmetric-key authentication. For every method, the \emph{ipsec} structure always specifies the protocol mode, the security protocol and the SAs' lifetime.  Instead, the \glspl{SPI}, algorithm and cryptographic keys are specified in different ways, depending on the specific key establishment method. That is, if the \gls{DP} method is used, this set of information are explicitly provided. Otherwise, that is \gls{IKE} is used as \gls{KMP}, this set of information is not explicitly provided, but rather negotiated and established when the Client and RS performs \gls{IKE}.

The choice of the particular method to use should be driven by the capabilities of the Client and RS, as well as by the policies and infrastructure used in the specific use case for provisioning and managing key material. In particular, the DP method is extremely efficient and hence preferable for very constrained devices, as the Client and RS do not take the explicit burden to establish an \gls{SA} pair. However, it does not provide strict assurances in terms of perfect forward secrecy. On the other hand, the two methods based on IKEv2 do provide perfect forward secrecy, as a native feature of the IKEv2 protocol. However, this requires the Client and RS to perform a full establishment of their \gls{SA} pair through IKEv2, with a consequent considerable commitment in terms of resources. The particular choice among IKEv2 symmetric-key and asymmetric-key authentication method really depends on the key infrastructure of the specific use case. While Certificate-based public keys are typically more cumbersome to handle and process, they are often preferable to pre-shared keys that do result in more efficient processing while at the same in more complicated provisioning and management operations. In the following, we provide more details about the three key establishment methods.

\noindent
\textbf{1) Direct Provisioning (DP)}. In this method, the \gls{SA} pair is pre-defined by the \gls{AS}. That is, SA-RS and SA-C are specified in the access token and in the \gls{RS} Information of the Access Token Response that the \gls{AS} sends to the Client. Note that the \gls{AS} cannot guarantee the uniqueness of the \emph{SPI\_SA\_C} identifier at the \gls{RS}, and of the \emph{SPI\_SA\_RS} identifier at the Client. In order to address possible collisions with a previously defined SPI, the \gls{AS} generates \emph{SPI\_SA\_C} and \emph{SPI\_SA\_RS} as random values. By doing so, the probability of a collision to occur is at most $2^{-32}$ for 32-bit long \gls{SPI}s.

In case a collision occurs at the \gls{RS}, i.e. the \gls{RS} receives an access token with a \emph{SPI\_SA\_C} value already used by another \gls{SA}, the \gls{RS} replies to the Client with an error message and aborts the setup of the IPsec channel. In network scenario scenarios where such additional overhead is not affordable, it is possible to reserve in advance a pool of \gls{SPI} values intended to be used only with the \gls{DP} method. This pool is exclusively managed by the \gls{AS}. Then, when an IPsec channel is closed and the related pair of SAs become stale, the \gls{RS} asks the \gls{AS} to restore the SPI of that SA-C as available.

Instead, in case a collision occurs at the Client, i.e. the Client receives a \emph{SPI\_SA\_RS} value already used by other \gls{SA}, the Client sends a second Access Token Request to the \gls{AS}, asking for an updated access token. This token request also includes an \emph{ipsec} structure containing only the field \emph{SPI\_SA\_RS} specifying an available identifier to use. Then, the \gls{AS} replies with the corresponding Access Token and RS Information updated only as to the requested \emph{SPI\_SA\_RS}.

\noindent
\textbf{2) IKEv2 with symmetric-key authentication.}
This method uses the IKEv2 protocol to establish the SA pair between Client and RS, while providing mutual authentication through symmetric cryptography. The Client and RS run IKEv2 in symmetric mode, using a symmetric \gls{PSK} provided by \gls{AS} and bound to the access token as a \gls{PoP} key. The PSK is made available to the Client in the Access Token Response, and to the RS in the access token. If the Client is interacting with the \gls{RS} for the first time, the \gls{AS} includes also a unique key identifier of the PSK in the Access Token Response. Otherwise, the Client includes in the Access Token Request a key identifier pointing at a previously established \gls{PSK}.

\noindent
\textbf{3) IKEv2 with asymmetric-key authentication.}
This method uses the IKEv2 protocol to establish the SA pair between Client and RS, while providing mutual authentication through asymmetric cryptography. The Client and RS run IKEv2 in asymmetric mode, using their \gls{RPK} or \gls{CPK} bound to the access token as \gls{PoP} keys. The RS's RPK/CPK is made available to the Client in the Access Token Response, while the Client's RPK/CPK is made available to the RS in access token. Similarly to the previous method, if the Client is interacting with the \gls{AS} for the first time, it includes its \gls{RPK} or \gls{CPK} in the Access Token Request. Otherwise, the Client includes a key identifier linked to its own \gls{RPK} or \gls{CPK}, which is already available at the \gls{AS}.

\subsection{Protocol Description}
\begin{figure}[t]
\centering
\includegraphics[width=\columnwidth]{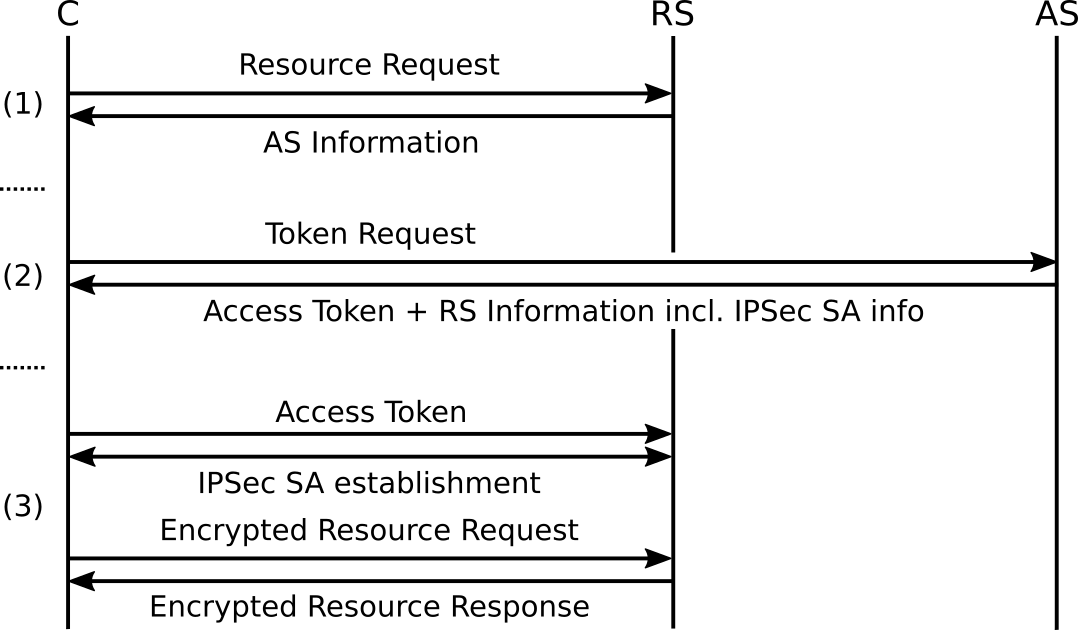}
\caption{IPsec profile message exchange (adapted from \cite{aragon-ace-ipsec-profile-01}).}%
\label{fig:ipsec_prot}
\end{figure}

In this section, we describe the message exchanges occuring in the \gls{ACE} framework, in the presence of our \gls{IPsec} profile. Intuitively, the workflow consists of three phases, as shown in Figure~\ref{fig:ipsec_prot}.

\noindent
\textbf{Phase (I) - Unauthorized Client to Resource Server}. During this phase, the Client can retrieve information necessary to contact the \gls{AS}, unless already available. In particular, the Client sends an \emph{unauthorized request} to the \gls{RS}, which formally denies the request and replies by indicating the associated \gls{AS} to contact for obtaining an access token.

\noindent
\textbf{Phase (II) - Client to Authorization Server}. During this phase, the Client sends an Access Token Request to the \emph{/token} endpoint at the \gls{AS}, indicating the resource of interest at the RS and the access \emph{scope}, i.e. the intended operations on such resource. Then, the AS processes the Access Token Request and verifies that the Client is allowed to access the specified protected resource at the RS. In such a case, the AS replies with an access token and the RS information as part of the Access Token Response. In particular, the access token (RS information) includes parameters and key material intended for the RS (the Client) to set up an IPsec as a pair of SAs. 

The exact information to exchange between the Client and the \gls{AS} depends on the \gls{SA} establishment method \ref{subsec:methods}. Unlike the \gls{DP} method, the alternative ones require the Client and the \gls{RS} to establish the \gls{SA} pair by running \gls{IKE}. To this end, the \gls{AS} indicates the specific \gls{KMP} to use in the \emph{kmp} field of the access token and of the RS Information. Specifically, \emph{kmp} is set to \emph{"ikev2"} to signal the use of the \gls{IKE} protocol. Provided that the involved parties have the necessary support, it is possible to use and specify a different key management protocol. Note that the \gls{AS} is aware of the Client's and \gls{RS}'s capabilities as well as of \gls{RS}'s preferred and supported communication settings \cite{ietf-ace-oauth-authz-07}. Therefore, the \gls{AS} is able to set the security and network Parameters for the \gls{SA} pair consistently with that Client-RS pair.

\noindent
\textbf{Phase (III) - Client to Resource Server} In this phase, the Client posts the access token to the \emph{/authz-info} endpoint at the AS, through a POST \gls{CoAP} message. 
Then, the Client and the \gls{RS} set up the \gls{SA} pair and the IPsec channel, based on the establishment method signalled by the AS. In particular:

\emph{a}) The {\gls{DP}} method is signalled by the presence of the \emph{ipsec} structure, while the \emph{"COSE\_Key"} field is not present.\footnote{The \emph{"COSE\_Key"} is a \gls{CBOR} object protected by using \gls{COSE}, and containing information about the used key material, such as key type, key identifier, and the actual cryptographic key.}.

\emph{b}) A symmetric-key authenticated establishment is signalled by including a \emph{"COSE\_Key"} object with the key type parameter \emph{"kty"} set to \emph{"Symmetric"}, and by indicating the usage of \gls{IKE} with the \emph{kmp} field set to \emph{"ikev2"}.

\emph{c}) An asymmetric-key authenticated establishment is indicated by including a \emph{"COSE\_Key"} object with the key type parameter \emph{"kty"} indicating the usage of asymmetric cryptography, e.g. \emph{"EC"}, by and indicating the usage of \gls{IKE} with the \emph{kmp} field set to \emph{"ikev2"}.

In case the \gls{DP} method is used, the Client and the \gls{RS} already have all the information to start the IPsec channel, and do not need to explicitly interact with each other. Instead, if any of the authenticated establishment methods is used, the Client and the \gls{RS} perform an actual SA pair establishment through \gls{IKE} according to the authentication mode indicated by the \emph{"kty"} field. In the following, we describe how the client and Client and RS finalize/setup the IPsec channel, given the specific establishment method.

\emph{a}) \emph{Direct Provisioning}. The Client derives all the necessary key material from the \emph{"seed"} field of the \emph{"ipsec"} structure in the RS Information. The Client uses the \textit{seed} to perform the a key derivation algorithm as in \gls{IKE}~\cite{rfc7296}. Upon correct submission and successful verification of the access token at the \emph{/authz-info} endpoint, the \gls{RS} performs the same key derivation process. The \gls{RS} replies to the Client over the IPsec channel, according to what specified in SA-RS. Thereafter, any further communication performed during the lifetime of the access token occurs over the IPsec channel defined by the SA pair.

\emph{b}) \emph{Authenticated SA Establishment using IKEv2}.
\label{par:authenticated_establishment_using_ikev2}
the Client and the \gls{RS} run the \gls{IKE} protocol, and use the key material in the respectively received \emph{"COSE\_key"} object in order to achieve mutual authentication. In particular, the Client posts the access token to the \emph{/authz-info} endpoint to the \gls{RS}, which sends back the first IKEv2 message IKE\_SA\_INIT to acknowledge the correct reception of the access token. Depending on the type of key used as \gls{PoPK}, i.e. symmetric or asymmetric, the \gls{IKE} protocol is executed in the corresponding mode \cite{rfc7296}, i.e. \gls{PSK}, \gls{CPK} or \gls{RPK}, with no modifications. If the \gls{IKE} execution is successfully completed, the Client and the \gls{RS} agree on key material, parameters and algorithms used to enforce the IPsec channel.

\section{Performance Evaluation and Discussion} %
\label{sec:performance_evaluation}
\begin{figure}[t]
\centering
\includegraphics[width=0.9\columnwidth]{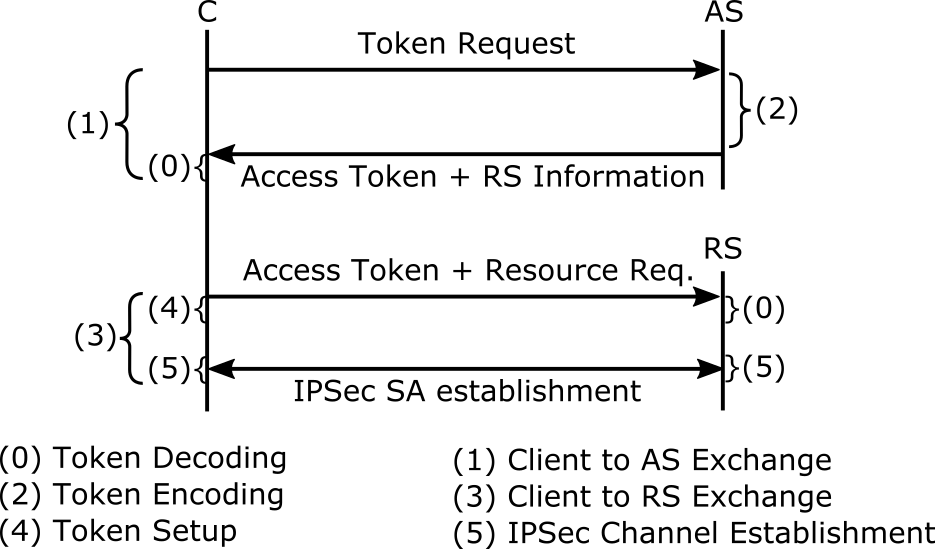}
\caption{Evaluation setup}
\label{fig:eval_image}
\end{figure}

In this section we present the performance evaluation of the different key establishment methods for the \gls{ACE} IPsec Framework, described in Section~\ref{s:protocol_overview}. 
The implementation is written for Contiki OS \cite{contiki} based on existing libraries and protocol implementations \cite{mkovatsc-2011-iotech-coap}. In particular, our code was evaluated on the Zolertia Firefly platform, equipped with the CC2538 radio chipset, 32 kB of RAM and 512 kB of flash ROM \cite{zolertia}. A device of this class supports a power supply from two AA (AAA) batteries, each of which typically provides an energy content of $9.36$ ($5.07$) KJ. 

Our implementation leverage on hardware-based cryptography and utilizes the following algorithms: AES-CCM* to secure the IEEE 802.15.4 link layer and the \gls{COSE} objects; AES-128 to provide confidentiality for \gls{IPsec} and \gls{IKE} with an 8-bytes long \gls{IV}; ECC-DH with 256-bit Random ECP Group \cite{rfc5114}, SHA-2 and a \gls{HMAC} based on SHA-256 for the authenticated exchange of \gls{IKE} \cite{rfc4894,rfc4868}.  IPsec and \gls{IKE} traffic is encrypted using ESP in transport mode.
The scenario to be evaluated is the following: a Client requests access a protected resource stored in a constrained \gls{RS}. The authentication and authorization of this request are delegated to the \gls{AS}. 
In our experimental setup the \gls{AS} is as well a resource-constrained device and performs routing related activities. Namely, it is set as the root of the \gls{DAG} and performs housekeeping operations for \gls{RPL}\cite{rfc6550}.

We evaluate four setup configurations: a baseline configuration (\textbf{Base}), i.e. the ACE Framework w.o. the IPsec profile; and the three key establishment methods: \textbf{DP}, Establishment with symmetric key authentication (\textbf{IKE-PSK}) and Establishment with asymmetric key authentication (\textbf{IKE-CPK}) \cite{rfc7296}.

Our experimental results include: memory footprint, packet size and time and energy measurements. The latter measurements are evaluated as the shown in Figure~\ref{fig:eval_image} which depicts where time and energy measurements are performed. This measurements are labeled as follows: (0) for Access Token decoding; (1) for the Client to AS Exchange; (2) for Access Token and RS Information encoding; (3) for the Client to RS Exchange; (4) for the Access Token setup; and (5) for the IPsec channel establishment.

In Table \ref{tab:ace_packet_size} we provide the size of packets exchanged by our profile. The packet exchanges are labeled as in Figure ~\ref{fig:ace_prot}. This measurements reflect the size of the CoAP messages. The last row of Table \ref{tab:ace_packet_size} provides the size of the \emph{Access token}, which has a big influence on the message size.

\begin{figure}[t]
\centering
\includegraphics[width=.405\textwidth]{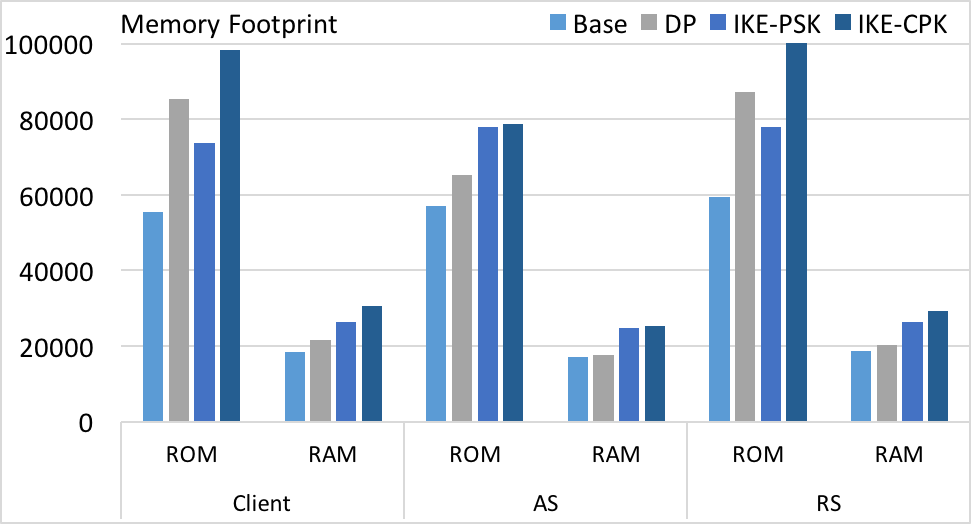}
\caption{Memory footprint comparison.}
\label{fig:memory}
\end{figure}
\begin{table}[t]
\centering
\caption{ACE Packet size and access token size in Bytes}
\label{tab:ace_packet_size}
\begin{tabular}{|c|c|c|c|c|}
\hline
\multicolumn{1}{|l|}{} & \multicolumn{1}{l|}{\textbf{Base}} & \multicolumn{1}{l|}{\textbf{DP}} & \multicolumn{1}{l|}{\textbf{IKE-PSK}} & \multicolumn{1}{l|}{\textbf{IKE-CPK}} \\ \hline
\textbf{(A)}                  & 10                             & 10                      & 10                           & 666                          \\ \cline{1-1}
\textbf{(B)}                  & 179                            & 401                     & 445                          & 1697                          \\ \cline{1-1}
\textbf{(C)}                  & 188                            & 299                     & 321                          & 947                          \\ \cline{1-1}
\textbf{(F)}                  & 23                             & 23                      & 23                           & 23                           \\ \cline{1-1}\hline
\emph{Access Token}             & 172                            & 283                     & 305                          & 931                          \\ \hline
\end{tabular}

\end{table}
In Figure \ref{fig:memory} we provide the Memory Footprint evaluation results for the different \gls{SA} establishment methods of the IPsec profile. We show the absolute value of ROM and RAM footprints for the setups configuration \textbf{Base}, \textbf{DP}, \textbf{IKE-PSK} and \textbf{IKE-CPK}.

Time measurements are collected using the system clock measured in system ticks. To convert our measurements to seconds the following formula is applied: \[time = \frac{sys\,clock}{ticks/second}\] 
Note that measurements (1) and (3) include network latency as round-trip time. Energy consumption measurements are divided into three contributions: energy spent at the CPU, energy spent in transmission state (TX) and the energy spent in reception state (RX).
To measure the energy consumed by the devices we use powertrace, a run-time power profiling mechanism which is part of Contiki. This tool has an accuracy of 94\% with an 0.6\% overhead \cite{powertrace}. The energy consumption out of the powertrace measurements is computed as follows:
    \[energy = \frac{powertrace\, value * current * voltage}{ticks/second}\]

For every setup configuration, 20 runs of the protocol were considered. We give average results for successful handshakes without packet loss. Note that wireless communication can be lossy in constrained environments with a loss rate typically increasing for larger packet sizes. In this case, the handshake duration as well as the energy consumption increase due to the retransmission of the packets.

\begin{figure}[t!]
\centering
\begin{subfigure}[b]{.45\columnwidth}
\includegraphics[width=\columnwidth]{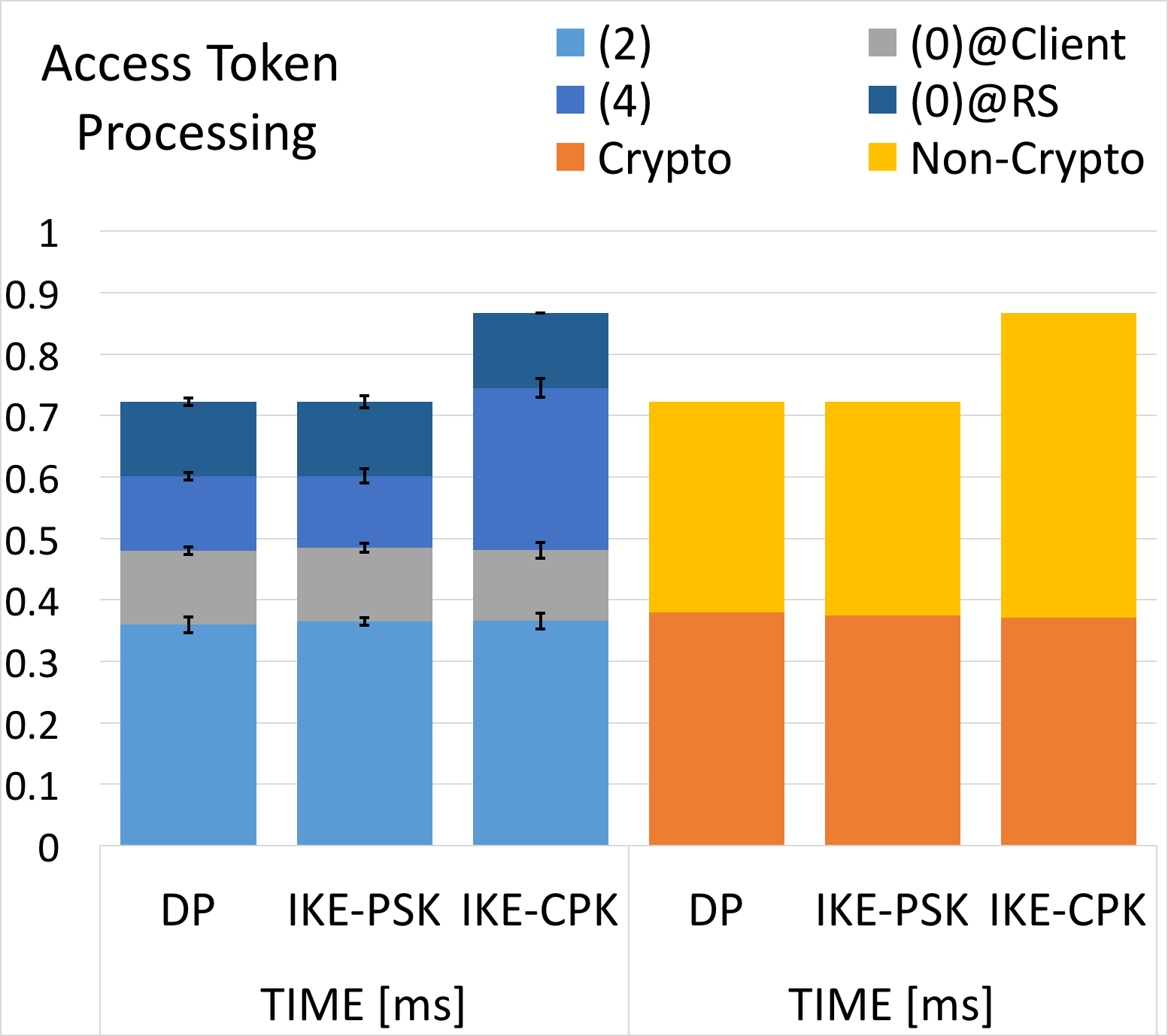}
\caption{Time measurements}
\label{fig:1_3_plot_1}
\end{subfigure}
\begin{subfigure}[b]{0.45\columnwidth}
\includegraphics[width=\columnwidth]{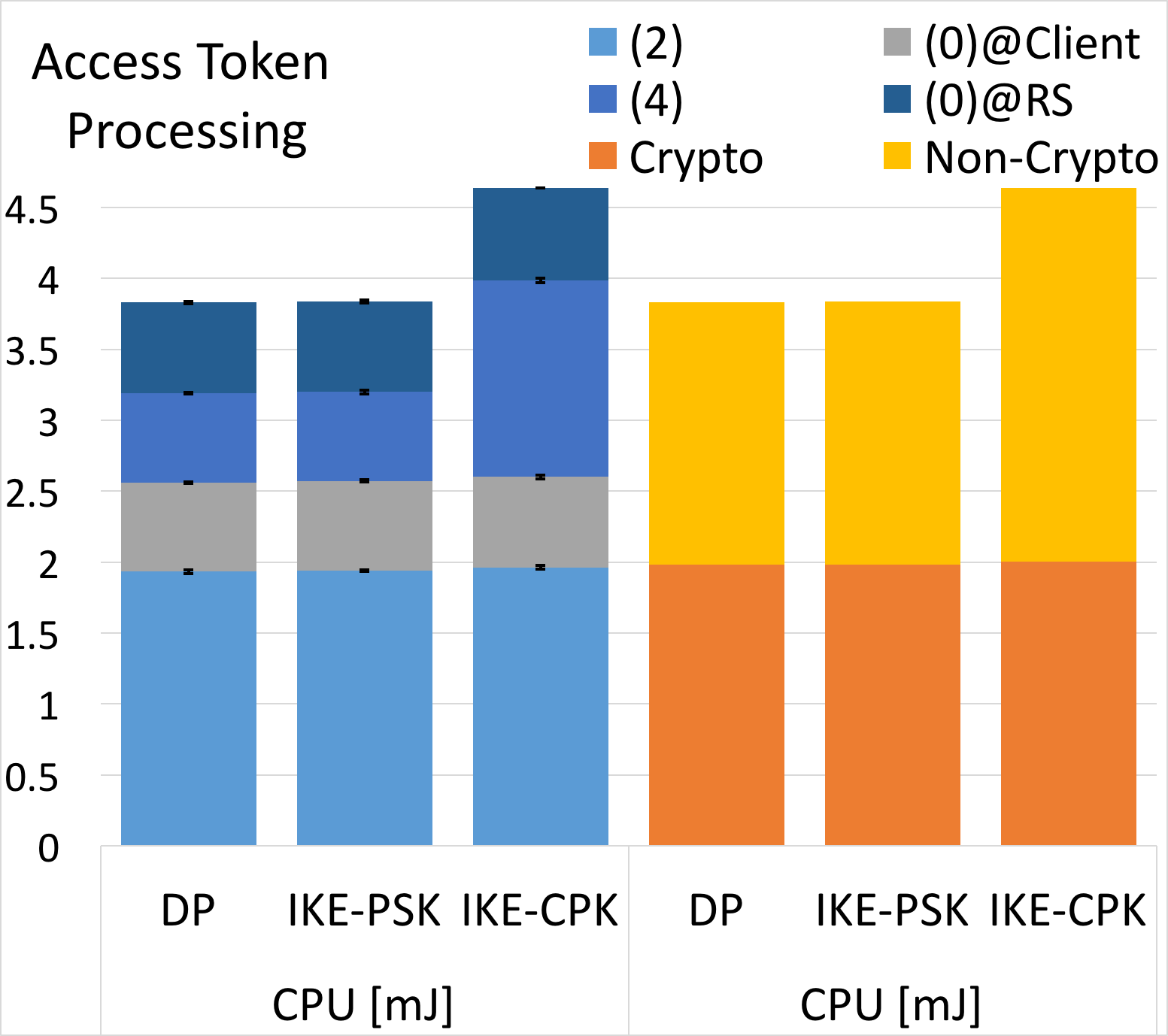}
\caption{Energy measurements}
\label{fig:1_3_plot_2}
\end{subfigure}
\caption{Token processing}
\label{fig:token}
\end{figure}

In Figure \ref{fig:token} we show the time and energy results of the access token processing. On the right side of the figure we depict the contribution of the four steps involving token processing, i.e. (0) Token decoding performed at the Client and the RS, (2) Token encoding and (4) Token Setup, as in Figure \ref{fig:eval_image}. On the left side, we categorized these operations in crypto- and non-crypto-related actions.

\begin{figure}[t!]
\centering
\begin{subfigure}[b]{.365\columnwidth}
\includegraphics[width=\columnwidth]{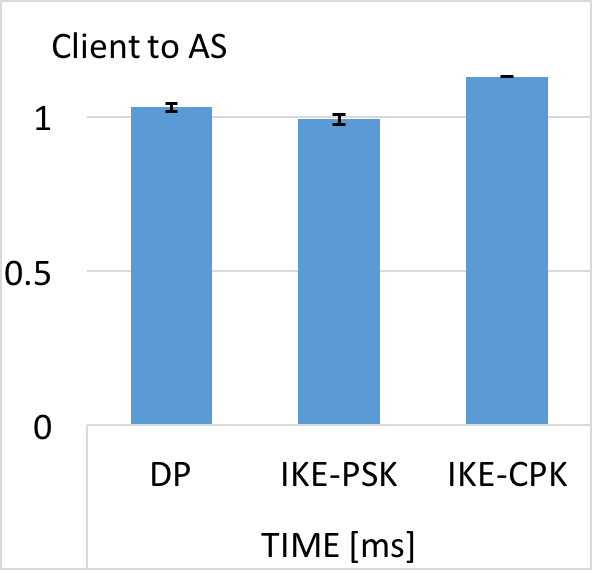}
\caption{Time measurements}
\label{fig:1_3_plot_3}
\end{subfigure}
\begin{subfigure}[b]{0.55\columnwidth}
\includegraphics[width=\columnwidth]{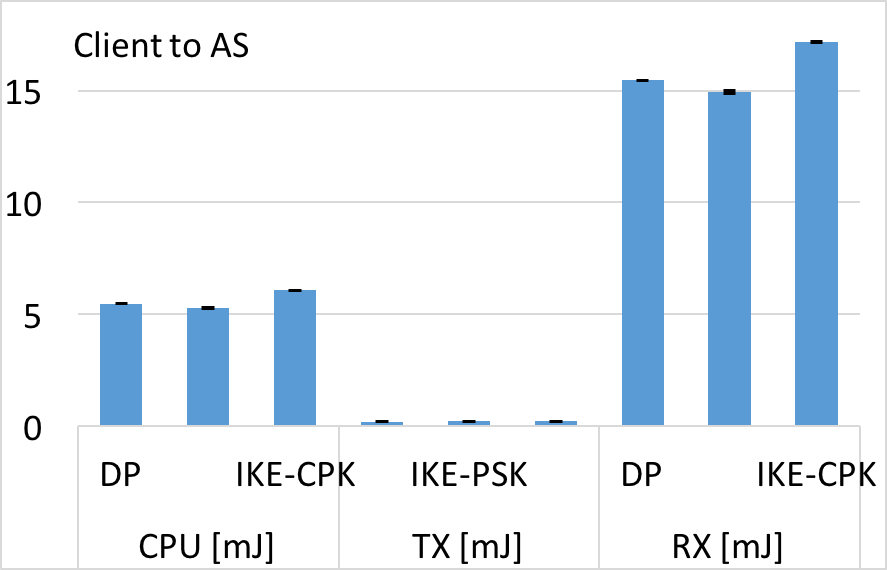}
\caption{Energy measurements}
\label{fig:1_3_plot_4}
\end{subfigure}
\caption{(1) Client to AS evaluation results}
\label{fig:c-as}
\end{figure}

The Client-to-AS message exchange results are shown in Figure~\ref{fig:c-as}. In this figure we can observe Client-to-AS network latency, processing time and energy consumption, i.e. measurement \textbf{(1)} in Figure~\ref{fig:eval_image}. A comparable performance disregarding the \gls{SA} establishment method can be observed. Namely, the Access Token Request/Response present a consistent behavior across the different setup configurations.

\begin{figure}[t!]
\centering
\begin{subfigure}[b]{.365\columnwidth}
\includegraphics[width=\columnwidth]{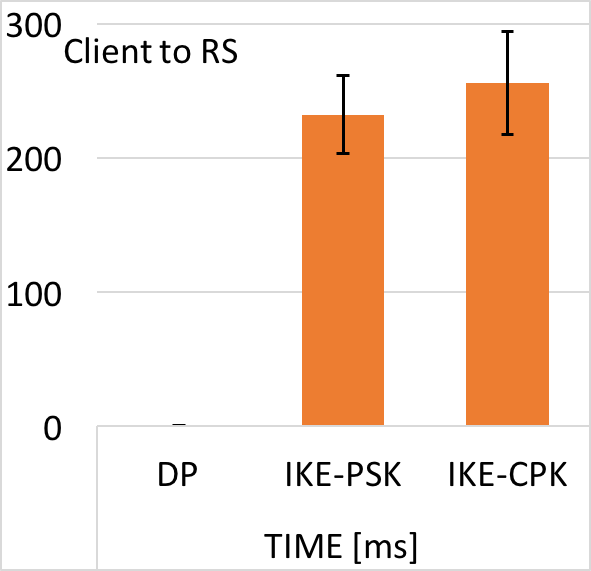}
\caption{Time measurements}
\label{fig:1_3_plot_5}
\end{subfigure}
\begin{subfigure}[b]{0.55\columnwidth}
\includegraphics[width=\columnwidth]{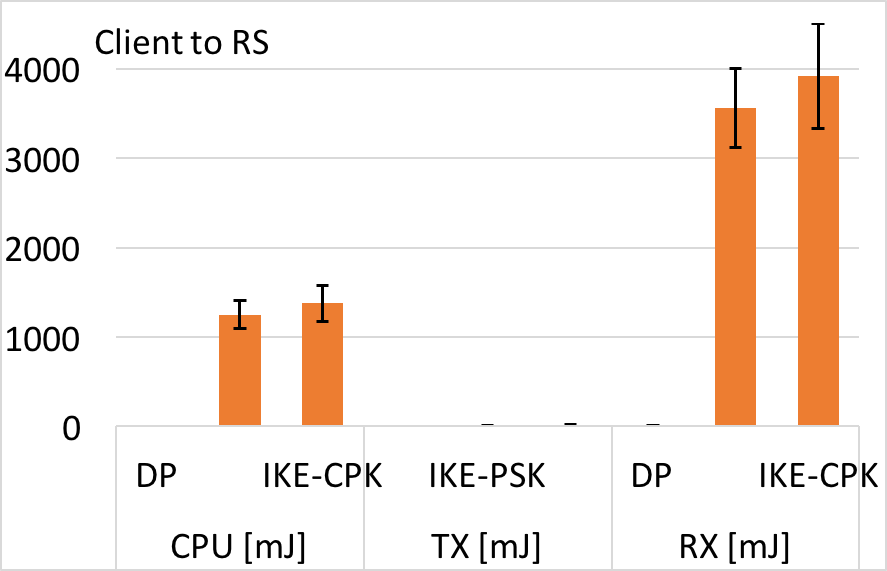}
\caption{Energy measurements}
\label{fig:1_3_plot_6}
\end{subfigure}
\caption{(3) Client to RS evaluation results}
\label{fig:c-rs}
\end{figure}

In Figure \ref{fig:c-rs} we show the Client-to-RS message exchange evaluation results, i.e. the measurement tag as (3) in Figure \ref{fig:eval_image}. We can observe that the results for the \gls{IKE}-based methods are comparable with the  \textbf{IKE-CPK}, showing a slightly bigger energy consumption. At the same time, \textbf{DP} time and energy results are notably lower than the IKE-based key establishment methods. The total energy spent in a \textbf{DP} establishment for (3) is on average 15 mJ, and the exchange is done in less than 1 ms on average.

\begin{figure}[t!]
\centering
\begin{subfigure}[b]{.365\columnwidth}
\includegraphics[width=\columnwidth]{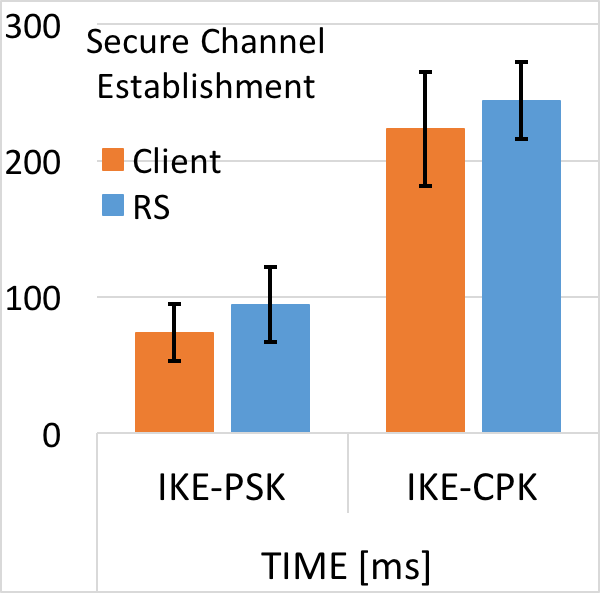}
\caption{Time measurements}
\label{fig:1_3_plot_7}
\end{subfigure}
\begin{subfigure}[b]{0.55\columnwidth}
\includegraphics[width=\columnwidth]{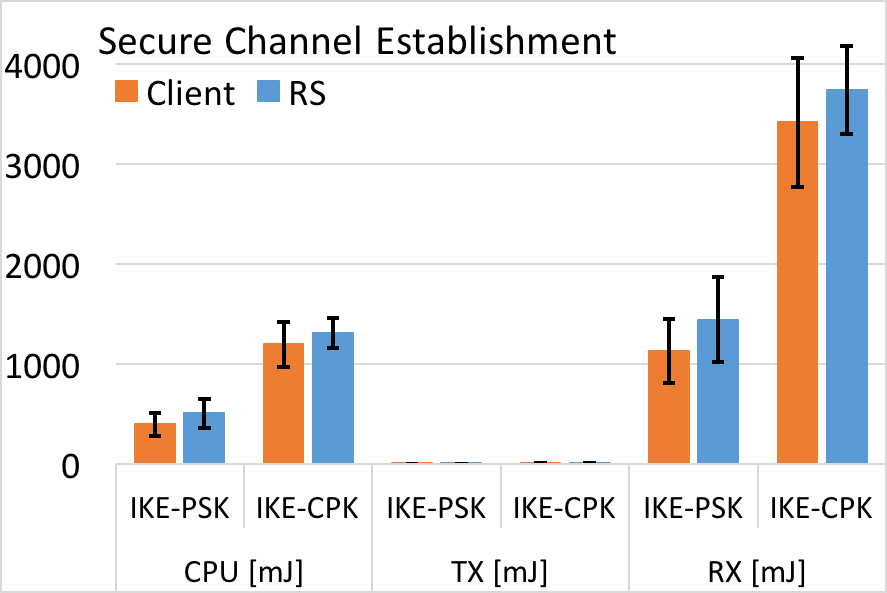}
\caption{Energy measurements}
\label{fig:1_3_plot_8}
\end{subfigure}
\caption{(5) Secure Channel Establishment}
\label{fig:5_plot}
\end{figure}

The evaluation results of the the establishment of a secure channel between RS and the Client are shown in Figure ~\ref{fig:5_plot}. Note that only IKE-based establishments perform an IPsec SA establishment, since in the \textbf{DP} method the IPsec SAs are provided by the AS. The Client and the RS perform similarly during (5), as in Figure~\ref{fig:eval_image}. This result is aligned with the symmetric nature of the IPsec protocol, since both ends of the communication play a similar role, unlike protocols like \gls{DTLS} where there are a client and a server role with different responsibilities.  However, within this symmetry, it is noticeable that for the \gls{RS}, (5) takes longer than for the Client. The aforementioned difference reflects the fact that the \gls{RS} is the initiator party of the IPsec channel establishment.

We can see that the energy spent in transmission state, label as TX in Figures \ref{fig:c-as}, \ref{fig:c-rs} and \ref{fig:5_plot} appears negligible when compared with the energy spent at the CPU or in receiving state, labeled as RX in the aforementioned figures. On the other hand, RX measurements in the aforementioned figures represent a significant share of the energy spend during a protocol run. This behavior is due to the fact that the reception state is always set to \textit{on} in our resource-constrained devices. Energy optimization techniques such as \gls{RDC}, specified and benchmarked in ~\cite{article}, are out of the scope of this work

\section{Conclusion}
\label{s:conclusion}
This paper has presented our novel ACE IPsec profile for authentication and authorization in the IoT. Our profile enables the scalable and flexible establishment of IPsec communication channels between Clients and Resource Servers, while contextually enforcing fine-grained access control from the ACE framework. In particular, IPsec Security Associations can be either directly provided to Client and Resource Server, or established though the standard IKEv2 key management protocol. We have implemented the IPsec profile for the Contiki OS, and carried out an experimental performance evaluation, considering resource-constrained IoT devices of the Zolertia Firefly platform. Results show that, under different configurations and authentication modes, our ACE IPsec profile is affordable also in resource-constrained devices. Therefore, it is effectively deployable in IoT scenarios for successfully enforcing access control paired with IPsec-based secure communication. Future works will focus on implementing alternative profiles of ACE and comparing their performance in resource-constrained IoT settings.

\section*{Acknowledgment}
This work has been supported by the EIT-Digital High Impact Initiative ACTIVE; the EIT-Digital Master School; VINNOVA CEBOT; VINNOVA Eurostars SecureIoT; the DFG in the Collaborative Research Center CROSSING (project S1) and project C.1 within the RTG 2050 ``Privacy and Trust for Mobile Users''; the EIT-Digital Master School.

\bibliographystyle{IEEEtran}
\bibliography{IEEEabrv,references.bib}

\end{document}